# Assessing composition gradients in multifilamentary superconductors by means of magnetometry methods

T. Baumgartner,[1,*] J. Hecher,[1] J. Bernardi,[2] S. Pfeiffer,[2] C. Senatore,[3] and M. Eisterer[1]

[1]*Atominstitut, TU Wien, Stadionallee 2, 1020 Vienna, Austria*
[2]*USTEM, TU Wien, Wiedner Hauptstraße 8–10, 1040 Vienna, Austria*
[3]*Département de Physique de la Matière Quantique, Université de Genève, Quai Ernest-Ansermet 24, 1211 Geneva, Switzerland*

We present two magnetometry-based methods suitable for assessing gradients in the critical temperature and hence the composition of multifilamentary superconductors: AC magnetometry and Scanning Hall Probe Microscopy. The novelty of the former technique lies in the iterative evaluation procedure we developed, whereas the strength of the latter is the direct visualization of the temperature dependent penetration of a magnetic field into the superconductor. Using the example of a PIT $Nb_3Sn$ wire, we demonstrate the application of these techniques, and compare the respective results to each other and to EDX measurements of the Sn distribution within the sub-elements of the wire.

## I. INTRODUCTION

While superconducting materials are usually characterized in terms of their critical temperature $T_c$, the reality is that technical superconductors often exhibit a $T_c$ distribution rather than a single value.[1–3] This is particularly relevant in superconductors in which diffusion reactions play a role during manufacturing, leading to compositional gradients. Since one usually strives to produce a homogeneous superconducting phase, knowledge of the spatial distribution of the critical temperature – and hence of the composition, if the dependence of $T_c$ on stoichiometry is known – can be a valuable feedback for the production process. In this paper we describe two magnetometry-based methods for probing spatial $T_c$ distributions. One of them, which we will refer to as the AC magnetometry method in the following, allows a quick characterization of multifilamentary superconducting wires, as long as certain conditions are met. The other, which makes use of Scanning Hall Probe Microscopy, may not be as convenient as the first, but it has fewer restrictions, and offers a more complete characterization than the AC magnetometry method.

In the following we will focus on $Nb_3Sn$ multifilamentary conductors, in particular PIT (powder-in-tube) wires, for two reasons. First, in this material the critical temperature exhibits a strong dependence on stoichiometry, and such wires are prone to compositional gradients due to their production method, which makes them an ideal model system.[4,5] And second, research on $Nb_3Sn$ is currently experiencing a renaissance driven by the desire to equip the Future Circular Collider with 16 T dipoles based on this material.[6]

The stoichiometry of $Nb_3Sn$ is usually expressed in terms of the atomic Sn content $\beta$, casting the chemical composition in the form $Nb_{1-\beta}Sn_\beta$. The superconducting A-15 phase exists in the range $\beta \approx 0.18-0.255$, corresponding to a $T_c$ range extending from approx. 6 K to somewhat above 18 K. Even more relevant from an application-oriented point of view is that the upper critical field $B_{c2}$ is also strongly affected by stoichiometry, peaking at a Sn content of approx. 24.5%. $Nb_3Sn$ for technical applications is usually a ternary system ($(Nb_{1-x}Ta_x)_3Sn$ or $Nb_3(Sn_{1-y}Ti_y)$, although using other alloying elements as well as adding more than one dopant is also possible), in which the substitution of Nb or Sn by other elements serves to enhance $B_{c2}$ by increasing the normal-state resistivity. Detailed reviews of the superconducting properties of $Nb_3Sn$ and how they vary with composition can be found in Ref. 4 and 7.

The production of PIT $Nb_3Sn$ wires starts with tubes made of Nb (or Nb alloyed with e.g. Ta), which are filled with a powder consisting mostly of $NbSn_2$. These tubes are optionally sheathed with a diffusion barrier, and stacked inside a Cu matrix. Mechanical deformation of this billet eventually yields the final unreacted wire. The actual A-15 phase is obtained by applying a heat treatment, leading to a solid state diffusion reaction in which – after the formation of some intermediate phases – tubular $Nb_3Sn$ sub-elements are formed.[5] Modeling this process is a complex matter, but recently progress was made in this field.[8]

This diffusion process results in a Sn concentration gradient inside of each sub-element, pointing from the outer edge towards the core. This gradient is sensitive to the reaction temperature and duration, and, given a sensible choice for the heat treatment, takes on values of typically $0.1-0.4$ at.%/µm.[5,9,10] Another characteristic feature of PIT $Nb_3Sn$ wires is their twofold grain morphology: the sub-elements consist of fine grains ($\sim 150$ nm) on the outside, whereas large grains ($\sim 1-2$ µm) dominate near the core. It was found that these two grain types exhibit significantly different $T_c$ and $B_{c2}$ distributions.[11] The large grains are undesired, since they are generally believed to contribute very little to current transport under conditions of technological relevance, since grain boundaries are known to be the primary flux pinning centers in $Nb_3Sn$.[12]

* tbaumgartner@ati.ac.at

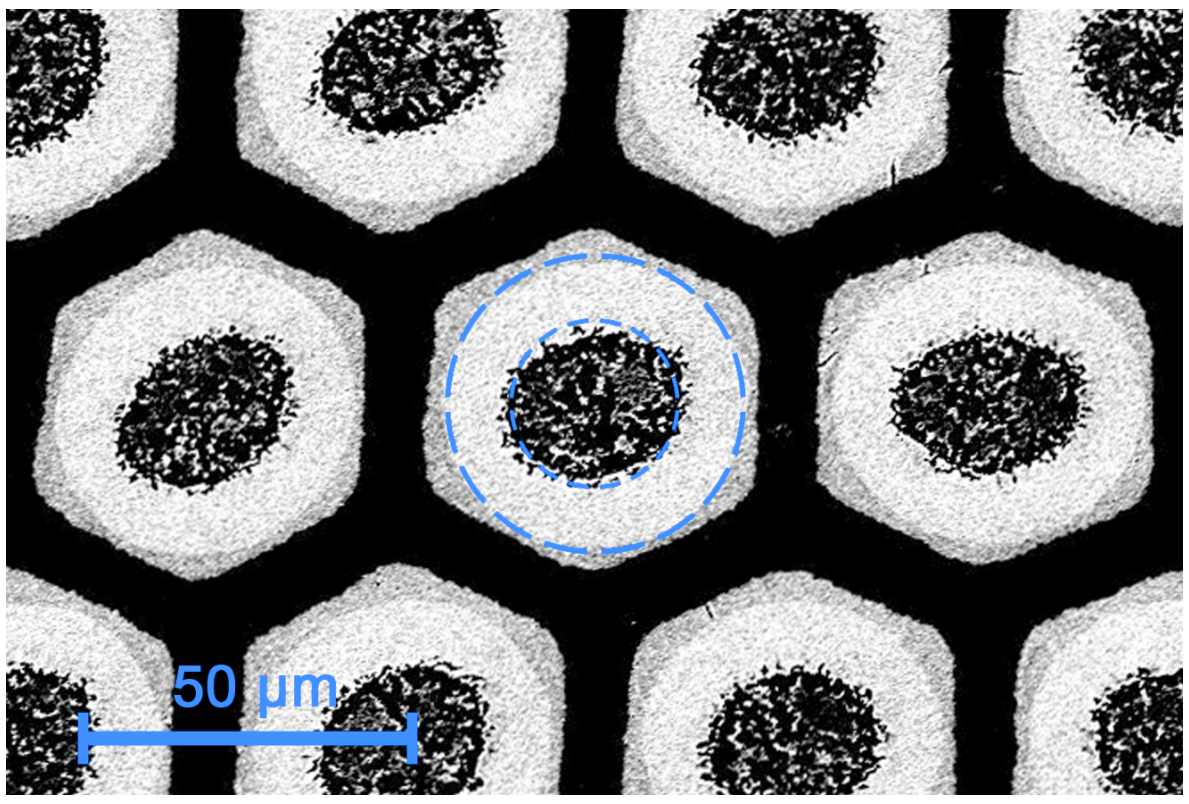

FIG. 1. SEM image of sub-elements as seen in a cross section of the examined wire. The A-15 area (light gray), the barrier (darker shade of gray), the core (black with bright speckles), and the copper matrix (black) can be clearly discerned. The average dimensions of the A-15 area are indicated by dashed circles superimposed on the sub-element in the center.

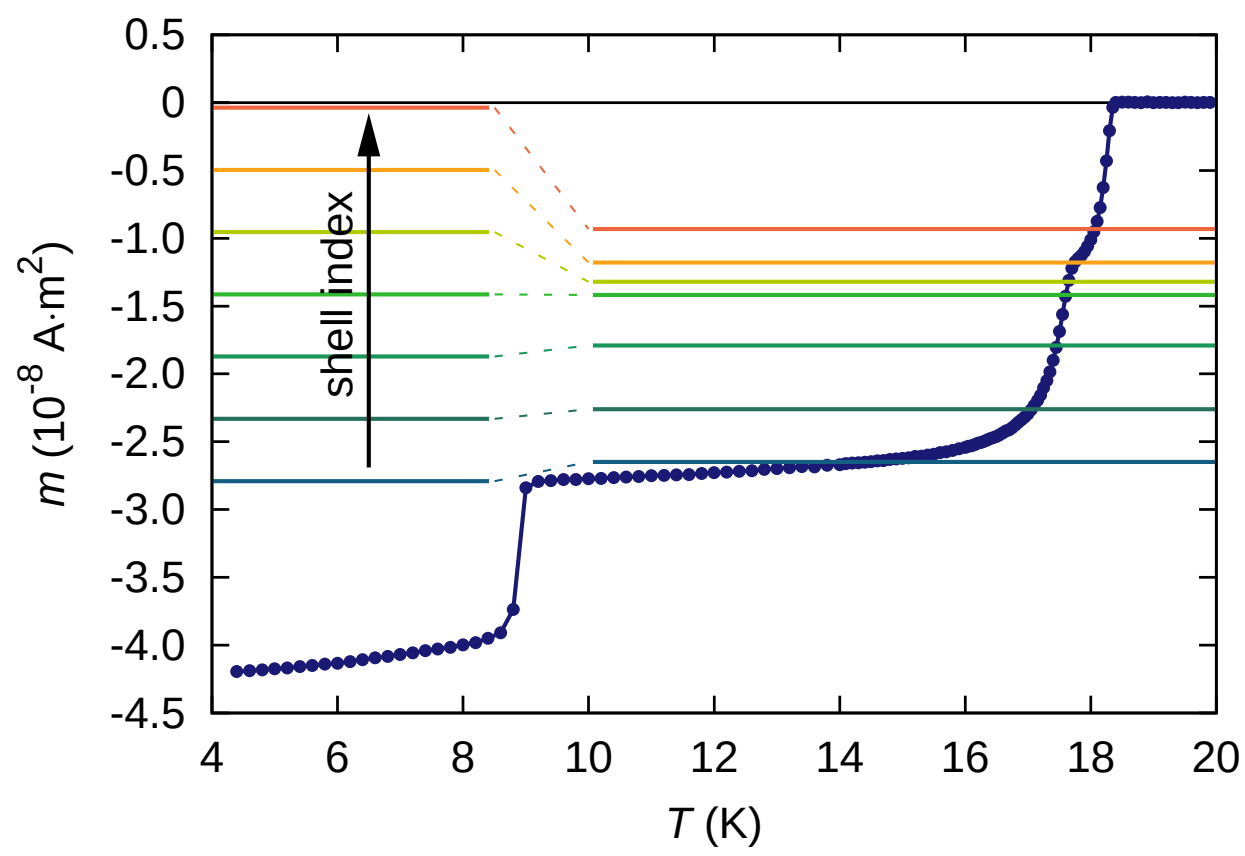

FIG. 2. Magnetic moment of a 4 mm long wire sample as a function of temperature. The measurement was performed by applying a small AC magnetic field parallel to the wire axis. The horizontal lines represent the computed moments the evaluation algorithm iterates to obtain the shell $T_c$ values (left: initial moments, right: final set of moments; dashed lines serve as a guide to the eye).

## II. EXPERIMENTAL DETAILS

The wire examined in this work is a state-of-the-art Ta-alloyed PIT $Nb_3Sn$ wire produced by Bruker EAS. It contains 192 sub-elements, some of which are depicted in the cross-sectional SEM image shown in Figure 1 together with dashed circles indicating the inner and outer radius of the A-15 region of a sub-element. These radii were determined by averaging over pixel counts obtained from 10 sub-elements, assuming ideal circular cross sections, and are relevant to the evaluation procedure described in section II A.

The spatial variation of the Sn concentration within sub-elements of the wire was examined by means of Energy-Dispersive X-ray (EDX) analysis using a Quanta 250 FEG Scanning Electron Microscope (SEM). Since this is an established technique for analyzing chemical compositions, we use the EDX data to assess the quality of the results obtained from the magnetometry-based techniques described in the following. This comparison is presented in section III.

### A. AC Magnetometry Method

Due to the facts that sub-elements of PIT $Nb_3Sn$ wires have a higher Sn content on the inside, and that the critical temperature of $Nb_3Sn$ increases monotonically with $\beta$, the sub-elements are "magnetically transparent". That means with increasing temperature a magnetic field applied parallel to the wire will penetrate a sub-element from the barrier towards the core, thus allowing to probe the radial $T_c$ distribution.[1]

Our measurements were carried out using a Quantum Design MPMS XL SQUID magnetometer. A 4 mm long sample of the wire was placed inside the magnetometer, and an alternating magnetic field with an amplitude of 30 μT and a frequency of 33 Hz was applied parallel to the wire using the AC option of the system. The magnetic moment arising from Meißner screening was then measured as a function of temperature, which was increased in small steps from $\sim$5 K to $\sim$20 K. The thus obtained data are shown in Figure 2. At low temperatures the Nb barriers around the sub-elements are screening the applied field, as evidenced by the jump at approx. 9 K. Above this value the magnitude of the signal slowly decreases with increasing temperature, suggesting a relatively good homogeneity. A kink can be found around 17.8 K, which corresponds to the temperature at which the fine grains are already in the normal state, whereas the large grains are still superconducting. The signal finally disappears at 18.4 K, where the superconducting volume in the sample reaches zero.

While the measurement principle is similar to that used by Hawes *et al.* quite some time ago (cf. Ref. 1), our evaluation procedure is much more elaborate. It is based on the following assumptions:

- The sub-elements inside the sample are parallel tubes with circular cross sections.
- All sub-elements are identical in terms of geometry and composition.
- Each sub-element exhibits a monotonic radial Sn gradient with the highest value on the inside, but no azimuthal Sn content variation.

If these assumptions are valid, it is sufficient to simulate a single sub-element by changing its radial $T_c$ distribution until the computed magnetic moment as a function of temperature matches the experimental data divided by the number of sub-elements. This is done by performing the following steps (a flowchart of the operations carried out by the evaluation software is provided as supplementary material):

1. Sub-divide the sub-element into $N$ concentric shells of equal thickness, and assign an initial $T_c$ value to each shell.
2. Outer shell loop: increment shell index $i$, i.e. go through all shells from the outside of the sub-element towards the inside.
    3. Inner shell loop: increment shell index $k \in [i+1, N]$, i.e. go through all shells deeper inside than shell $i$.
        4. Calculate the magnetic penetration depth of each shell assuming the temperature equals the $T_c$ value of shell $i$.
        5. Compute the magnetic moment resulting from Meißner screening in the shells using the penetration depth calculated for each of them.
    6. Adjust the $T_c$ value of shell $i$ by finding the temperature at which the computed magnetic moment intersects the experimentally determined $m(T)$ curve.
7. Continue the iteration process (step 2 *et seq.*) until the changes are negligible.

The initial $T_c$ values of the shells are chosen using the same procedure employed in step 6. This is accomplished using magnetic moments equally spaced between the value directly after the transition of the Nb barrier and the minimum non-zero absolute value, as indicated by the horizontal lines on the left-hand side of Figure 2.

The algorithm uses the two-fluid model expression for the temperature dependence of the magnetic penetration depth:

$$\lambda(t) = \lambda(0)\frac{1}{\sqrt{(1-t^4)}}\,, \quad t = \frac{T}{T_c} \tag{1}$$

We chose the literature value of 124 nm for the zero-temperature penetration depth $\lambda(0)$.[13] This assumption is certainly a source of error, since it does not account for the (to our knowledge unknown) dependence of $\lambda$ on the Sn content. However, the results presented in section III suggest that the inaccuracy introduced by this simplification is minor. For reasons elucidated below, an effective penetration depth $\bar{\lambda}$ is computed for each shell, which allows for the introduction of an intra-granular variation of the Sn content. This is accomplished by averaging the superfluid density $n$ within a grain representative for a given shell over its volume $V$. Since $n \propto 1/\lambda^2$, we can write

$$\bar{\lambda} = \left[\frac{1}{V}\int \frac{1}{\lambda^2(\vec{r})}\mathrm{d}^3 r\right]^{-1/2}, \tag{2}$$

where $\lambda(\vec{r})$ is calculated using Equation 1 with a local $T_c(\vec{r})$. For the calculation of the local critical temperature as a function of the local Sn content $\beta$ we use the following equation published by Godeke in Ref. 7:

$$T_c(\beta) = \frac{T_c^{\min} - T_c^{\max}}{1 + \mathrm{e}^{(\beta-\beta_0)/\Delta\beta}} + T_c^{\max} \tag{3}$$

The values of the parameters in this equation given in the above cited publication were determined by fits to data obtained from binary $Nb_3Sn$, whereas the wire we examined is Ta-doped for $B_{c2}$ optimization. At this concentration the Ta can be expected to increase $T_c$ by $\sim$0.3 K, which is in good agreement with the maximum value of slightly more than 18.3 K appearing in our measurements.[14] Therefore, we obtained parameters more appropriate for our sample by fitting Equation 3 to data presented in Ref. 7 after adding 0.3 K to each $T_c$ value. We fixed $T_c^{\max} = 18.6$ K to ensure a reasonable value at the phase boundary ($T_c(\beta = 0.255) = 18.33$ K), and obtained $T_c^{\min} = 6.39$ K, $\beta_0 = 22.01 \cdot 10^{-2}$, $\Delta\beta = 9.21 \cdot 10^{-3}$.

The effective penetration depth $\bar{\lambda}$ is used for computing the exponential decay of the magnetic field $H$ inside the sub-element:

$$H(x) = H(0)\,\mathrm{e}^{-x/\bar{\lambda}} \longrightarrow H_k = H_{k-1}\,\mathrm{e}^{-d/\bar{\lambda}_k} \tag{4}$$

In the above equation the subscripts denote shell indices, which increase from the outside towards the inside, and $d$ is the shell thickness. The current density $j_k$ in shell $k$ follows from Maxwell's equation $\nabla \times \vec{H} = \vec{j}$, and the shell current can be obtained by integrating over the cross section perpendicular to $\vec{j}$, yielding

$$\begin{aligned} I_k &= L\int_0^d j_k(x)\mathrm{d}x = -L\frac{H_{k-1}}{\bar{\lambda}_k}\int_0^d \mathrm{e}^{-x/\bar{\lambda}_k}\mathrm{d}x \\ &= LH_{k-1}\left(1 - \mathrm{e}^{-d/\bar{\lambda}_k}\right)\,, \end{aligned} \tag{5}$$

where $L$ is the sample length. With this shell current the total magnetic moment of the simulated sub-element can be expressed as a sum over the moments of the individual circular current loops, using the shell radii $r_k$:

$$m_i = \pi \sum_{k=i+1}^{N} r_k^2 I_k \tag{6}$$

This magnetic moment, which corresponds to Meißner screening by all shells with indices higher than $i$, is used to adjust the critical temperature of shell $i$. Intersections of the thus obtained magnetic moments with the experimental $m(T)$ curve are computed by means of linear interpolation between the data points to obtain new shell $T_c$ values. This process continues until the relative change of the magnetic moments between two consecutive iteration steps drops below a certain value. A criterion of 0.5% was found to be a reasonable choice, since smaller values do not have any discernible effect on the resulting $T_c$ distribution. The evolution of the computed magnetic moments corresponding to different values of $i$ in Equation 6 is indicated by the horizontal lines in Figure 2 (initial values on the left, moments after the final iteration step on the right). For the sake of clarity the number of shells was set to 7 for this plot, whereas a meaningful evaluation requires around 50 shells. Different reasonable choices for the magnetic moments from which the initial $T_c$ values are obtained (monotonic, covering the whole range of the transition) converge to the same distribution, however, the required number of iterations (typically around 20) may change.

Figure 3 shows results obtained from the examined wire sample using the algorithm described above. Three radial $T_c$ distributions, evaluated using different assumptions, are plotted as a function of the relative position between the innermost (position 0) and the outermost shell (position 1). The dash-dotted orange line serves to demonstrate that simplifying the evaluation by making the naive assumption of perfect shielding in each of the shells ($\lambda \to 0$, $m_i \propto r_i^2$) leads to a significantly different result. The dashed purple line shows the distribution obtained by taking the penetration depth into account, but without any intra-granular variation of the Sn content. It certainly looks plausible, and, contrary to the dash-dotted curve, it can withstand a comparison with EDX data (cf. section III). However, when a histogram is computed from this distribution, the result is a very narrow structure, as depicted in Figure 4 (dashed gray line).

A reliable way to obtain the actual $T_c$ histogram (but not the radial $T_c$ distribution, since the method does not offer spatial resolution) is the use of calorimetry combined with a deconvolution procedure.[15] Data from such a measurement performed on the very same sample we examined were available from previous work. As shown in Figure 4 by the red curve, whose individual data points are indicated by squares, the calorimetry histogram is much broader than the one we obtained. However, the two peaks corresponding to fine grains with lower $T_c$, and large grains with higher $T_c$, are located at the same positions as they are in the dashed gray curve. It is primarily the width of the fine grain peak which is significantly different.

At this point it is important to note that the length scale over which the respective experimental technique averages is the magnetic penetration depth $\lambda$ in the case of magnetometry, but it is the coherence length $\xi$ in calorimetry measurements. Hence, it stands to reason that the calorimetry method "sees" details inside of grains, which cannot be discerned by means of magnetometry. There is experimental evidence suggesting that $Nb_3Sn$ grains inside multifilamentary wires can exhibit a significant variation of the Sn content from the surface towards the center.[10,16] Therefore, we assume that intra-granular Sn gradients are responsible for the disagreement between the solid red and the dashed gray curves. To address this problem, we implemented the computation of an effective penetration depth which takes such variations into account (Equations (2) and (3)), and tested different gradients, assuming that the Sn content decreases linearly from the surface towards the center.

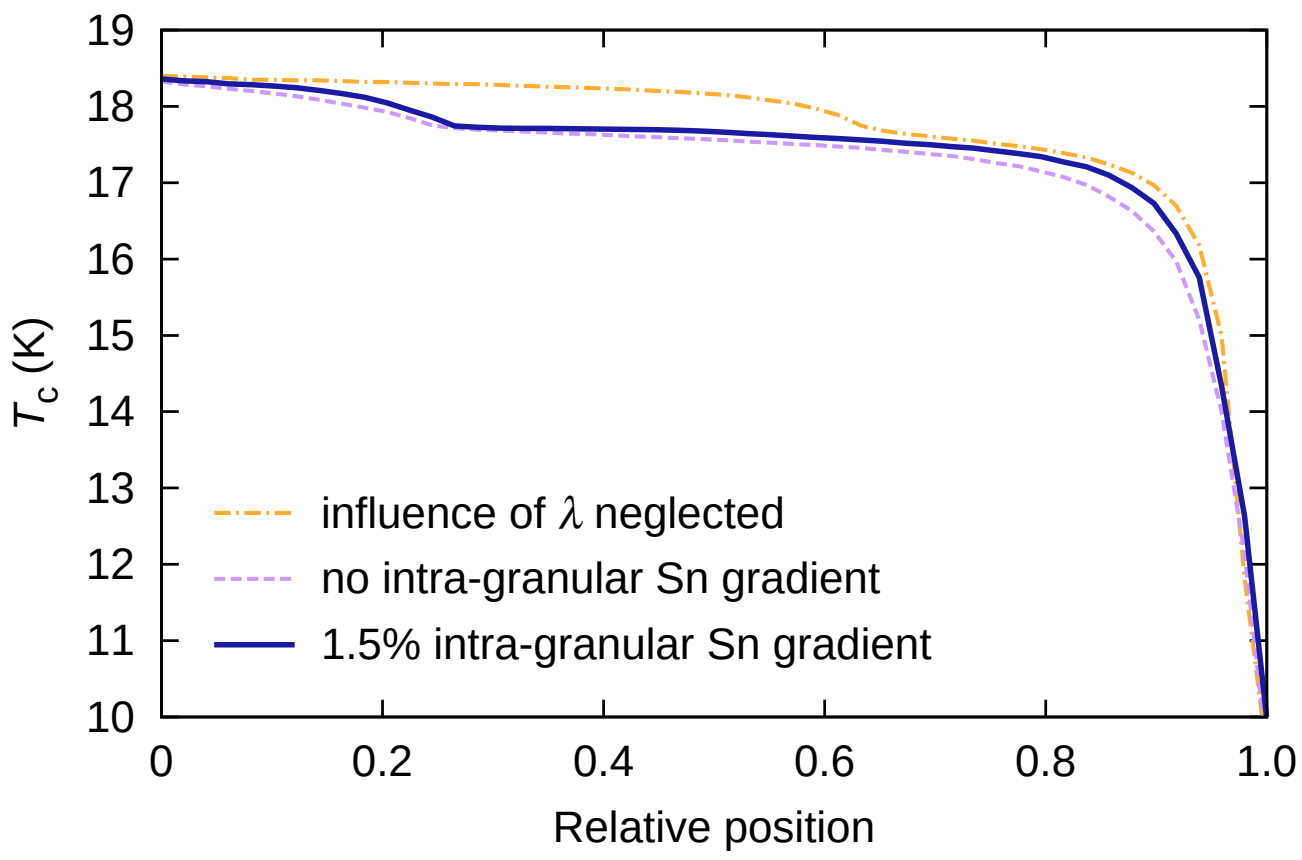

FIG. 3. Radial $T_c$ distribution within the sub-elements of the examined wire. The solid blue line, which was computed assuming an intra-granular Sn gradient of 1.5%, is the most reasonable result according to calorimetry data. The curves obtained without taking into account the penetration depth (dash-dotted orange line), and without an intra-granular gradient (dashed purple line), respectively, are shown for comparison.

A difference of 1.5% between grain surface and center was found to result in an acceptable agreement with calorimetry data. The corresponding $T_c$ distribution and histogram are shown by the solid blue lines in Figure 3 and Figure 4, respectively. A comparison of this result with the Sn distribution obtained from EDX is presented in section III.

### B. Scanning Hall Probe Microscopy

Scanning a Hall probe across the surface of a superconductor allows creating a magnetic field map, from which information on the current flow inside the sample can be inferred, e.g. by means of inversion of the Biot-Savart law. Our self-built Scanning Hall Probe Microscopy (SHPM) system offers a spatial resolution of $\sim 1\,\mu$m, and a scan range of $3 \times 3\,\text{mm}^2$. It is operated inside a helium flow cryostat, thus enabling measurements covering a wide temperature range. A detailed description of the set-up as well as examples of its performance can be found in Ref. 17.

In this work we used our SHPM system to directly visualize the penetration of a magnetic field into the sub-elements of a wire sample in the Meißner state. For that purpose a slice with a thickness of approx. 10 µm was cut from the sample previously examined using the AC magnetometry method. The surface facing the Hall probe was polished, then the disk was fixed on the sample platform of the SHPM system using vacuum grease, and cooled down in a flow cryostat. After the temperature had stabilized at the desired value, a magnetic field of 1 mT (or 0.3 mT in the case of the highest measurement temperature) was applied parallel to the sub-elements. A scan covering an area of $180 \times 180\,\mu\text{m}^2$ was then taken as close to the surface as possible ($\sim$1 µm distance). This procedure was performed over the same region of the wire at the temperature values 6 K, 10 K, 16 K, 17 K, and 17.8 K.

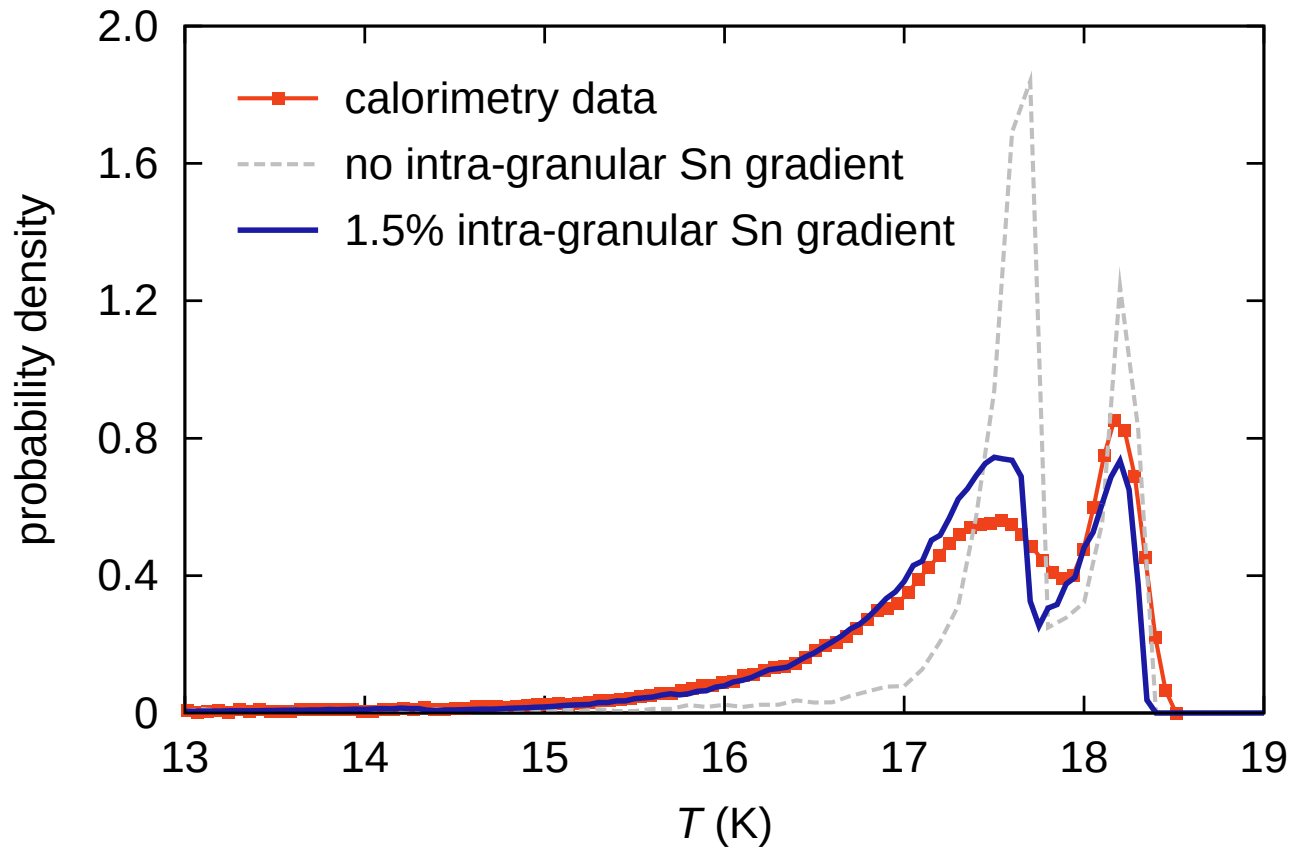

FIG. 4. Histograms of the $T_c$ distribution of the examined wire. The red line with squares representing data points was obtained by means of calorimetry using the method described in Ref. 15. The distribution derived from the AC magnetometry method yields a much too narrow histogram, if no Sn content variation within the grains is assumed (dashed gray line). Simulating an intra-granular Sn gradient of 1.5% (solid blue line) yields a reasonable agreement with the calorimetry data.

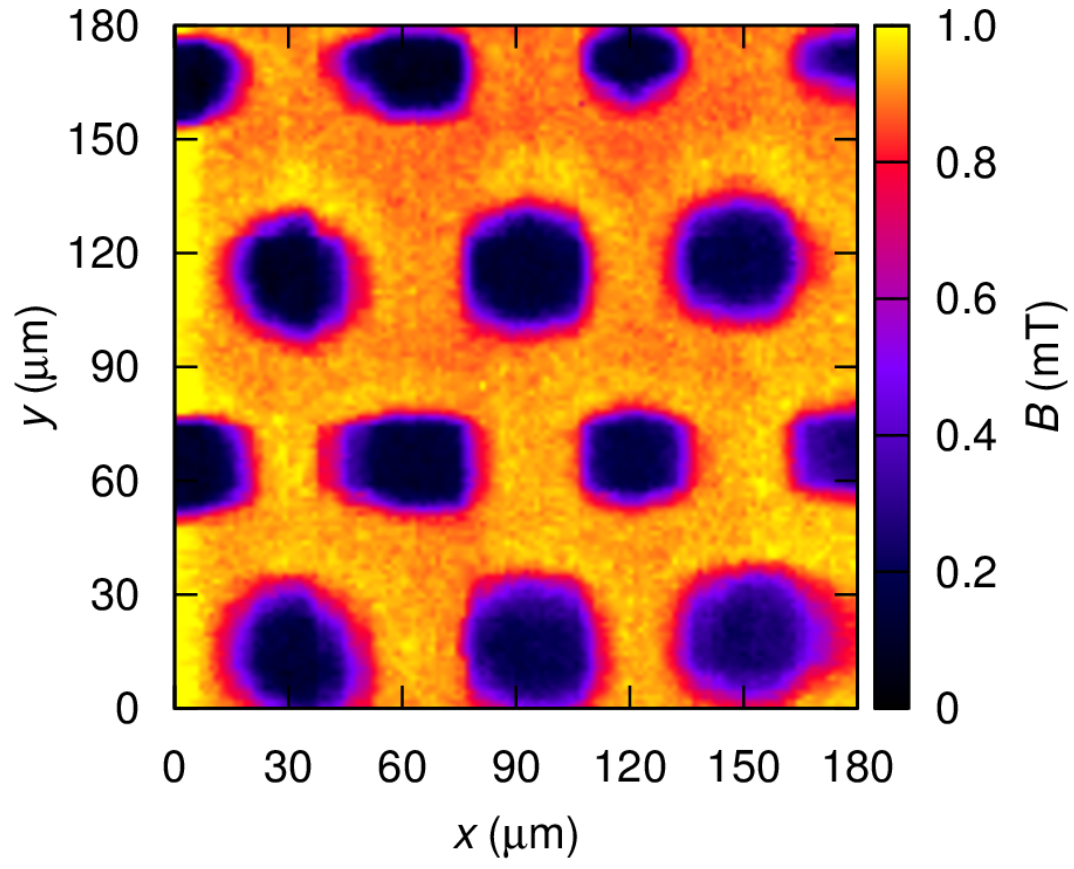

FIG. 5. Hall scan of the sub-elements inside a thin disk cut from the wire sample. The scan was taken at 17 K in an applied field of 1 mT.

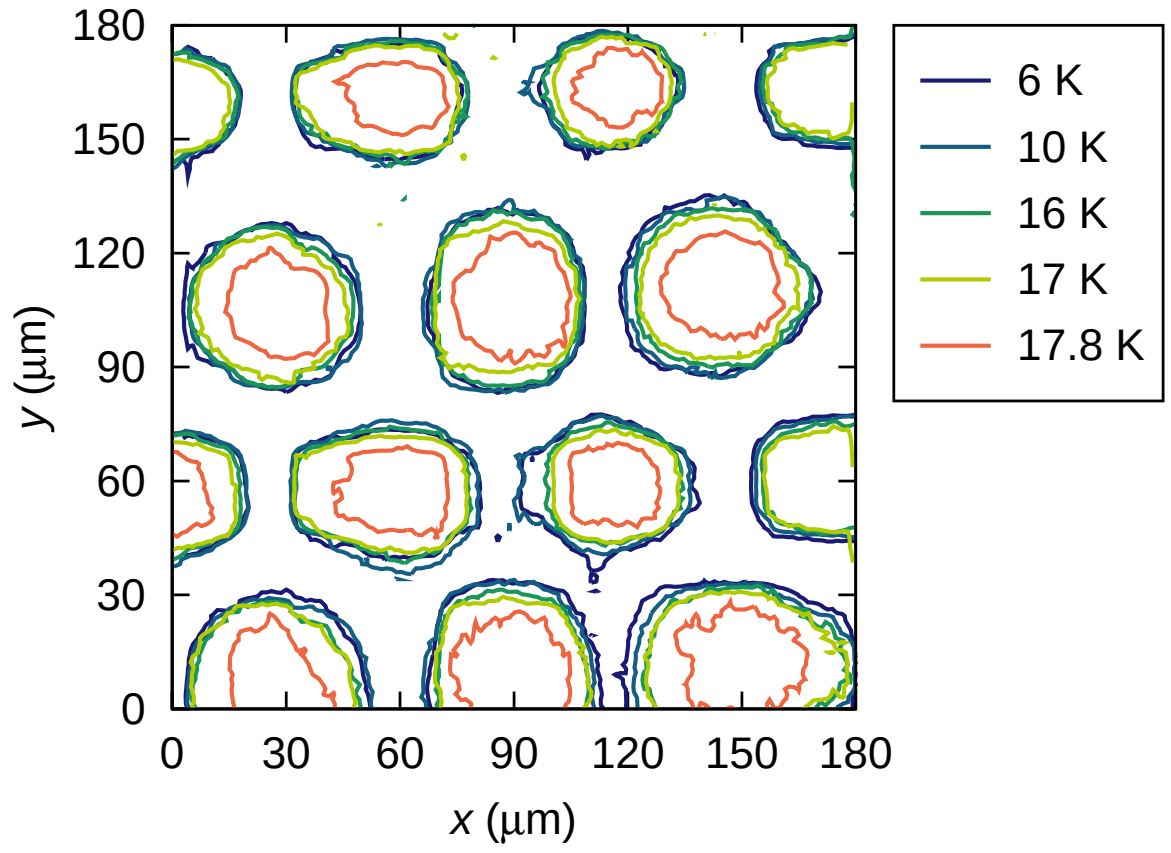

FIG. 6. Overlay of contour plots computed from Hall scans of the same region taken at five different temperatures. The lines correspond to the perimeter at which the local field has decayed to 80% of the applied field.

The scan taken at a temperature of 17 K is depicted in Figure 5. The shielded regions inside the individual sub-elements appear as dark spots whose hexagonal arrangement reflects the stacking matrix of the wire (cf. Figure 1). To visualize the shrinking of the shielded regions with increasing temperature, contour lines were computed from each of the scans, and overlaid, as shown in Figure 6. The lines indicate where the measured field equals 80% of the applied field at each temperature. Note that the increase of the penetration depth with increasing temperature is insufficient to cause the observed shrinking of the shielding perimeter. Hence, Figure 6 does indeed reveal the $T_c$ distribution within $Nb_3Sn$ sub-elements. A comparison to the results obtained from the AC magnetometry method and to EDX examinations is provided in section III.

While SHPM is a lot more laborious and time-consuming than the AC magnetometry method, it offers the advantage of being free from assumptions regarding the geometry of the superconductor. Structures with the highest $T_c$ on the outside, or even with non-monotonic $T_c$ distributions, can in principle also be assessed with SHPM by field-cooling the sample in an applied field greater than $H_{c1}(T)$, and measuring the remanent field after switching off the applied field. Another benefit lies in the ability to examine individual sub-elements, thus facilitating the investigation of differences between them (e.g. comparing the sub-elements close to the wire center to the ones near the surface, which are often deformed in the process of wire production).

## III. RESULTS AND DISCUSSION

As mentioned in the beginning of section II A, we performed EDX examinations on the wire sample in order to have a reference to which the magnetometry-based methods described in this work can be compared. To make this comparison possible, the radial $T_c$ distribution obtained from the AC magnetometry method (intra-granular Sn gradient of 1.5% assumed, as suggested by a comparison with calorimetry data) was converted to a Sn distribution using the inverse function of Equation 3. The contour plot assembled from Hall scan data was used to obtain the average radius of the shielded region (assuming circular geometry) inside a sub-element at different temperatures by measuring the areas enclosed by the respective contour lines. These radii were converted to relative positions between the inner and outer radius of the A-15 region, and the temperatures were converted to Sn content values. The value corresponding to $T = 6\,\mathrm{K}$ was set to $\beta = 0$, since it corresponds to shielding by the Nb barrier ($T_c \approx 9\,\mathrm{K}$). The coordinates of the EDX data points were transformed such that the A-15 phase boundaries on the inside and at the diffusion barrier are located at the relative positions 0, and 1, respectively.

Figure 7 shows a comparison of the results obtained from the AC magnetometry method, from SHPM, and from EDX. The EDX data are average values computed from measurements on five individual sub-elements, and the error bars indicate $\pm 2\sigma$, where $\sigma$ is the standard deviation. Evidently, the three sets of data are in rather good agreement. Morphology transitions (kinks in the Sn content) are observable both in the EDX data, and in the results obtained from the AC magnetometry method, the approximate positions being 0.35 in the former case, and 0.26 in the latter. The difference probably arises from variations among the sub-elements, as suggested by the large standard deviations of the EDX data in the range $\sim 0.25 - 0.35$. The slope $\mathrm{d}\beta/\mathrm{d}x$ computed from linear fits in the fine grain region of the data obtained from EDX and AC magnetometry is 0.11 at.%/µm in both cases, in agreement with results obtained by others in similar studies on PIT wires.[5,9,10]

In section II A we state that the introduction of an intra-granular Sn gradient is necessary to achieve agreement between the histogram obtained from the AC magnetometry method and the actual calorimetry data. One might therefore argue that the method is only useful if the intra-granular Sn gradient or the calorimetry histogram is known. We have to point out that this is not true, since the radial $T_c$ distribution is a lot less sensitive to the intra-granular Sn gradient than the histogram. Comparing the dashed purple curve to the solid blue curve in Figure 3 reveals that the difference between the $T_c$ distributions calculated with the assumption of zero intra-granular Sn gradient, and 1.5% gradient, respectively, is minor. This is a consequence of the relatively small impact of moderate intra-granular Sn gradients on the effective penetration depth (cf. Equation 2), which is visualized in Figure 8.

In Ref. 18 Tarantini *et al.* report the presence of disconnected grains with critical temperature values extending up to 18.7 K in a PIT wire very similar to the one studied in this work. The AC magnetometry method is not sensitive to grains which do not form a percolative path around the sub-element core (a "shell", as we named it in section II A), hence the absence of such high $T_c$ values in Figure 3 is not surprising. We did, however, not find any evidence for the existence of such grains in the calorimetry data, either.

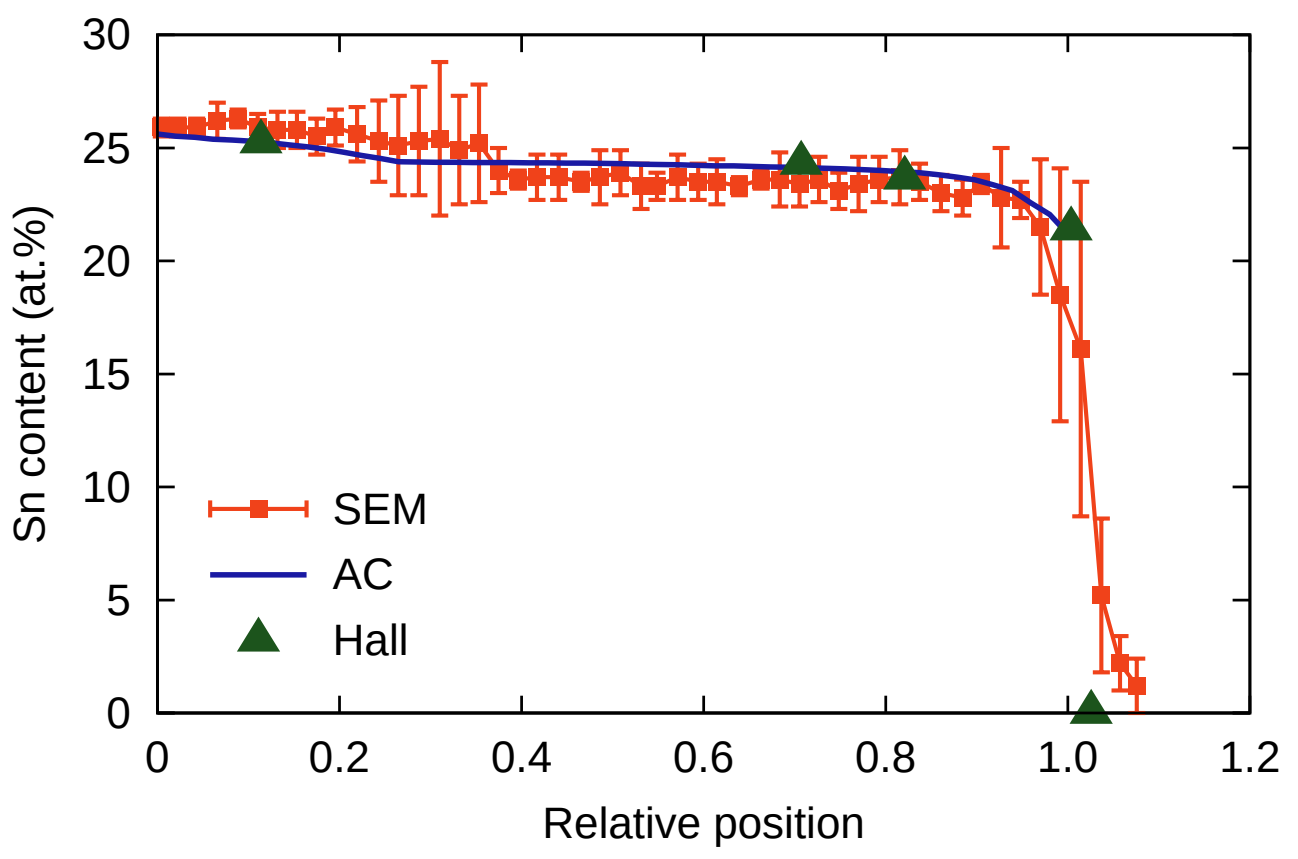

FIG. 7. Comparison of the radial Sn distributions obtained from EDX (solid red line with squares indicating the individual data points, and $\pm 2\sigma$ error bars), from the AC magnetometry method (solid blue line), and from SHPM (green triangles).

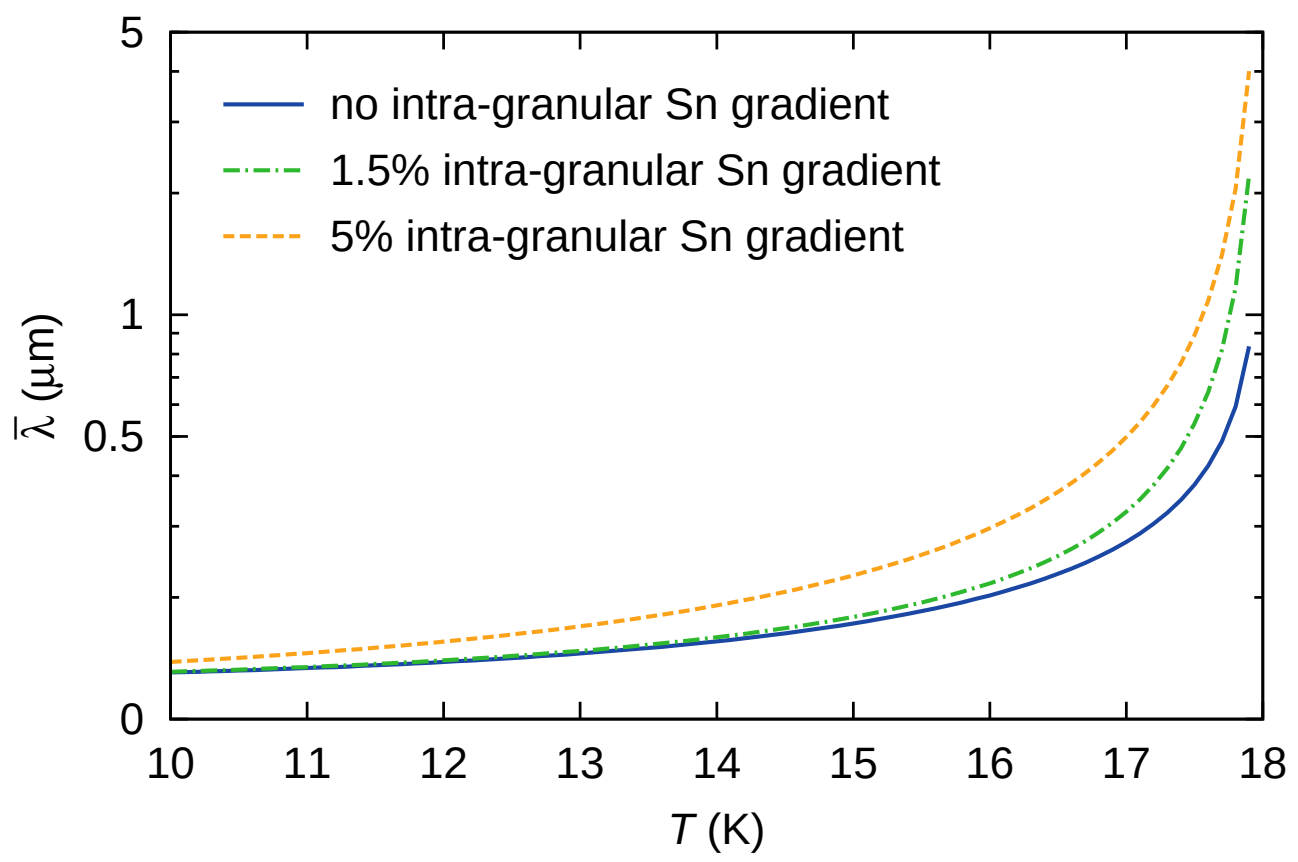

FIG. 8. Effective penetration depth as a function of temperature computed for different intra-granular Sn gradients and a shell $T_c$ (critical temperature on the surface of a representative grain) of 18 K.

## IV. CONCLUSIONS

In this paper we described two magnetometry-based methods for assessing composition gradients in multifilamentary superconductors, and demonstrated their suitability for examining PIT $Nb_3Sn$ wires by comparing the results with EDX measurements of the Sn distribution within the sub-elements. Although the experimental pro-cedure used in the AC magnetometry method is not new, our approach is novel in terms of the iterative evaluation procedure we developed. It is a relatively quick and easy technique for assessing $T_c$ gradients, which appears to produce reliable results. To our knowledge the other method we discussed, Scanning Hall Probe Microscopy, had hitherto not been used to assess the homogeneity of multifilamentary wires. While it is a lot slower and more challenging than the use of a SQUID magnetometer, it offers the direct visualization of magnetic screening, without the need for assumptions regarding the geometry.

PIT wires are simple in that their sub-elements exhibit a monotonic radial Sn gradient, whereas Internal Tin wires can develop two Sn concentration gradients: a global gradient stemming from the outward diffusion of the Sn core, and a local gradient arising from Sn diffusion into the individual filaments inside the bundles.[19] It is currently unclear whether the AC magnetometry method is suitable for such wires. In our opinion it could be, if the individual filaments inside the bundles form a well-connected A-15 region during the heat treatment, and if the local Sn variation is sufficiently small in magnitude or significant only on a length scale much smaller than $\lambda$.

The intra-granular Sn gradient we inferred from the discrepancy between the histograms obtained from the AC magnetometry method and from calorimetry should – although backed up by literature – be experimentally confirmed. We plan to investigate the Sn distribution within individual grains of the wire discussed in this work by means of EDX measurements using a Scanning Transmission Electron Microscope. In addition to that, examinations involving different wire types, and an investigation of the effects of A-15 inhomogeneities on the critical current density are envisaged. A better understanding of the diffusion and phase formation processes in $Nb_3Sn$ can probably lead to advances in wire performance. We are confident that the characterization techniques discussed in this work can contribute to this development.

## ACKNOWLEDGMENTS

We would like to acknowledge the support by CERN, who funded research at Atominstitut and provided samples within our fruitful collaborations (contract numbers K-1726/TE, KE1943/TE, and KE2254/TE). Furthermore, we would like to thank Jakob Gruber for taking care of the sample preparation at USTEM.

[1] C. D. Hawes, P. J. Lee, and D. C. Larbalestier, IEEE Trans. Appl. Supercond. **10**, 988 (2000).
[2] C. Senatore and R. Flükiger, Supercond. Sci. Technol. **22**, 095016 (2009).
[3] S. X. Dou, O. Shcherbakova, W. K. Yeoh, J. H. Kim, S. Soltanian, X. L. Wang, C. Senatore, R. Flükiger, M. Dhalle, O. Husnjak, and E. Babic, Phys. Rev. Lett. **98**, 097002 (2007).
[4] R. Flükiger, D. Uglietti, C. Senatore, and F. Buta, Cryogenics **48**, 293 (2008).
[5] A. Godeke, A. den Ouden, A. Nijhuis, and H. H. J. ten Kate, Cryogenics **48**, 308 (2008).
[6] "FCC Week 2016," April 11–15, 2016, Rome, Italy. 'Storify' web link: Superconductivity and the FCC.
[7] A. Godeke, Supercond. Sci. Technol. **19**, R68 (2006).
[8] X. Xu and M. D. Sumption, Sci. Rep. **6**, 19096 (2016).
[9] C. D. Hawes, P. J. Lee, and D. C. Larbalestier, Supercond. Sci. Technol. **19**, S27 (2006).
[10] M. Cantoni, C. Scheuerlein, P.-Y. Pfirter, F. de Borman, J. Rossen, G. Arnau, L. Oberli, and P. Lee, J. Appl. Phys.: Conference Series **234**, 022005 (6 pp.) (2010).
[11] C. Senatore and R. Flükiger, Appl. Phys. Lett. **102**, 012601 (2013).
[12] R. M. Scanlan, W. A. Fietz, and E. F. Koch, J. Appl. Phys. **46**, 2244 (1975).
[13] A. Godeke, *Performance Boundaries in $Nb_3Sn$ Superconductors*, Ph.D. thesis, University of Twente (2005), ISBN: 90-365-2224-2.
[14] M. Suenaga, D. O. Welch, R. L. Sabatini, O. F. Kammerer, and S. Okuda, J. Appl. Phys. **59**, 840 (1986).
[15] C. Senatore, D. Uglietti, V. Abächerli, A. Junod, and R. Flükiger, IEEE Trans. Appl. Supercond. **17**, 2611 (2007).
[16] D. Rodrigues Jr., C. L. H. Thieme, D. G. Pinatti, and S. Foner, IEEE Trans. Appl. Supercond. **5**, 1607 (1995).
[17] J. Hecher, *Current Transport in Polycrystalline Iron Based Superconductors*, Ph.D. thesis, TU Wien (2016), http://www.ub.tuwien.ac.at/diss/AC10774721.pdf.
[18] C. Tarantini, C. Segal, Z. H. Sung, P. J. Lee, L. Oberli, A. Ballarino, L. Bottura, and D. C. Larbalestier, Supercond. Sci. Technol. **28**, 095001 (2015).
[19] C. Tarantini, P. J. Lee, N. Craig, A. Ghosh, and D. C. Larbalestier, Supercond. Sci. Technol. **27**, 065013 (2014).